\documentclass[runningheads,anonymous]{llncs}
\usepackage[T1]{fontenc}

\usepackage{amsmath,amssymb,amsfonts}
\usepackage{algpseudocode}
\usepackage[linesnumbered,ruled,vlined]{algorithm2e}
\usepackage{pifont}
\usepackage{graphicx}
\usepackage{textcomp}
\usepackage{xcolor}
\usepackage{pgf}
\usepackage{tikz}
\usepackage{color}
\usepackage{stfloats}
\usepackage{float}
\usepackage{subfig}
\usepackage{multirow}
\usepackage{diagbox}
\usepackage[hyphens]{url}
\usepackage{parallel}
\usepackage{wrapfig}
\usepackage{marvosym}
\usepackage{booktabs}
\usepackage{siunitx}
\usepackage{makecell}
\usepackage{threeparttable}
\usepackage{indentfirst}
\usepackage{bbm}
\usepackage{microtype}

\usepackage[normalem]{ulem}
\usepackage[sort,nocompress]{cite}
\usepackage{authblk}

\usepackage{booktabs}

\usepackage{comment}
\usepackage[bookmarks=true,breaklinks=true,colorlinks,linkcolor=black,citecolor=blue,urlcolor=black]{hyperref}
\usepackage{booktabs}
\usepackage{multirow}
\usepackage{diagbox}
\usepackage{threeparttable}
\usepackage{orcidlink}
\newcommand{\squishlist}{
   \begin{list}{$\bullet$}
    { \setlength{\itemsep}{0pt}      \setlength{\parsep}{0pt}
      \setlength{\topsep}{3pt}       \setlength{\partopsep}{0pt}
      \setlength{\listparindent}{-2pt}
      \setlength{\itemindent}{-5pt}
      \setlength{\leftmargin}{1em} \setlength{\labelwidth}{0em}
      \setlength{\labelsep}{0.5em} } }

\newcommand{\squishend}{
    \end{list}  }

\def\placeh{CoQMoE}
\begin{document}
\title{CoQMoE: Co-Designed Quantization and Computation Orchestration for Mixture-of-Experts Vision Transformer on FPGA}
\titlerunning{CoQMoE}
%
{\footnotesize
\author{Jiale Dong$^{*}$, Hao Wu$^{*}$, Zihao Wang, 
Wenqi Lou$^{\dagger}$ \orcidlink{0000-0002-2240-6672}, 
Zhendong Zheng,\\ Lei Gong, 
Chao Wang$^{\dagger}$ \orcidlink{0000-0002-9403-5575}, 
Xuehai Zhou \\}
}
\authorrunning{Dong et al.}

\institute{University of Science and Technology of China, Hefei, China \\
djl190011@mail.ustc.edu.cn \\
\{louwenqi, cswang\}@ustc.edu.cn \\}

%
\maketitle              
\let\thefootnote\relax
\footnotetext{
  \textsuperscript{*}Equal contribution. \textsuperscript{$^{\dagger}$}Corresponding authors.
}
\begin{abstract}
Vision Transformers (ViTs) exhibit superior performance in computer vision tasks but face deployment challenges on resource-constrained devices due to high computational/memory demands. While Mixture-of-Experts Vision Transformers (MoE-ViTs) mitigate this through a scalable architecture with sub-linear computational growth, their hardware implementation on FPGAs remains constrained by resource limitations. This paper proposes a novel accelerator for efficiently implementing 
quantized MoE models on FPGAs through two key innovations: (1) A dual-stage quantization scheme combining precision-preserving complex quantizers with hardware-friendly simplified quantizers via scale reparameterization, with only 0.28\% accuracy loss compared to full precision; (2) A resource-aware accelerator architecture featuring latency-optimized streaming attention kernels and reusable linear operators, effectively balancing performance and resource consumption. Experimental results demonstrate that our accelerator achieves nearly 155 frames per second, a 5.35$\times$ improvement in throughput, and over 80\% energy reduction compared to state-of-the-art (SOTA) FPGA MoE accelerators, while maintaining <1\% accuracy loss across vision benchmarks. Our implementation is available at https://github.com/DJ000011/CoQMoE.
\end{abstract}

\section{Introduction} \label{sec:intro}

The Vision Transformer (ViT) has garnered significant attention for its outstanding performance in various computer vision tasks~\cite{dosovitskiy2020image,chen2022dearkd,yun2024shvit,wang2022via,glvlsi24_qin}. Building on the Mixture-of-Experts (MoE) architecture, MoE-ViT extends the original ViT by scaling model size without proportional increases in computational complexity, thereby enhancing multitasking capabilities and emerging as a focal point of recent research~\cite{aaai22_moevit,chen2023adamv,liu2021swin,aspdac24_swat,kim2024monde,iscas25_ubimoe}. However, as the models grow in size, the rapid increase in the number of parameters introduces new challenges, particularly the need for more efficient computation strategies.

Current research on ViT and MoE-ViT optimizations predominantly falls into two categories: algorithmic improvements and hardware implementations. On the algorithmic front, a substantial body of work\cite{lin2022fqvit, li2023repq, tai2023tsptq, shi2024p} focuses on optimizing aspects such as quantization and sparsity, with a significant emphasis on quantization techniques aimed at reducing the computational footprint without compromising accuracy. From a hardware design perspective, FPGA-based ViT accelerators have become a topic of great interest. The architectures for these accelerators can generally be classified into two types: pipeline-based~\cite{wang2022via,guo2024hg} and reuse-based~\cite{dong2023heatvit,sarkar2023edge,tc_octcnn}. The reuse-based architecture typically instantiates a single processing element (PE) and relies on the host CPU for data exchange. In contrast, the pipeline-based architecture divides tasks into multiple sequential stages, with each stage executed by dedicated hardware, significantly improving frames per second (FPS) performance.

Despite the notable efficiency gains achieved by these accelerators for standard ViTs, they are often inadequate when applied to MoE-ViTs, due to the inherent differences in computational demands. 
\textit{Algorithmic Limitations}: Achieving a balance between hardware-friendly quantization algorithms and maintaining accuracy remains a significant challenge.  
Complex quantizers, \textit{e.g.}, FQViT~\cite{lin2022fqvit}, are proposed to minimize precision loss, which is challenging for FPGA implementation. Conversely, fully quantized methods improved hardware efficiency but incurred significant precision loss, \textit{e.g.}, RePQ~\cite{li2023repq}.
\textit{Hardware Limitations}: Adapting hardware architectures to accommodate the "dynamic loading" characteristic of MoE models is particularly difficult. For example, the pipeline-based architecture, which excels in sequential processing, conflicts with the random expert selection required in MoE models. This is evident in systems like HGpipe~\cite{guo2024hg}, where even relatively small models like ViT-Tiny cannot fully deploy all layers efficiently. Similarly, reuse architectures struggle with inefficiencies due to the need for separate instantiation of PEs for sparse and dense linear operations, further exacerbating performance bottlenecks.

In response to these challenges, our work aims to explore both the quantization opportunities and the hardware acceleration potential specific to MoE-ViTs. The key contributions of this paper are as follows:
\begin{itemize}
\item We propose a novel quantization algorithm that applies specialized quantizers to activations following LayerNorm and Softmax. Through reparameterization, the quantizers are transformed into a hardware-efficient version, enabling accelerator-friendly implementation while preserving accuracy.

\item We introduce a hybrid computation mode for hardware accelerators, including a streaming attention kernel designed to reduce latency and a unified sparse/dense linear kernel to enable efficient cross-layer computation in MoE and MLP modules. These components leverage computation reordering and broadcasting techniques to achieve O(1) off-chip memory access, regardless of the parallelization scale, thereby enhancing performance.

\item Our experimental evaluations demonstrate that the proposed approach outperforms the SOTA M$^3$ViT in key performance metrics, including speed and accuracy. Additionally, the proposed quantization algorithm and hardware accelerator prove effective not only for MoE-ViT models but also for accelerating standard ViT architectures.
\end{itemize}

\section{Background}
\subsection{Mixture-of-Experts Vision Transformer (MoE-ViT)}
\label{subsec:moe_vit}
\begin{figure}[t]
\centering
\includegraphics[width=.9\textwidth]{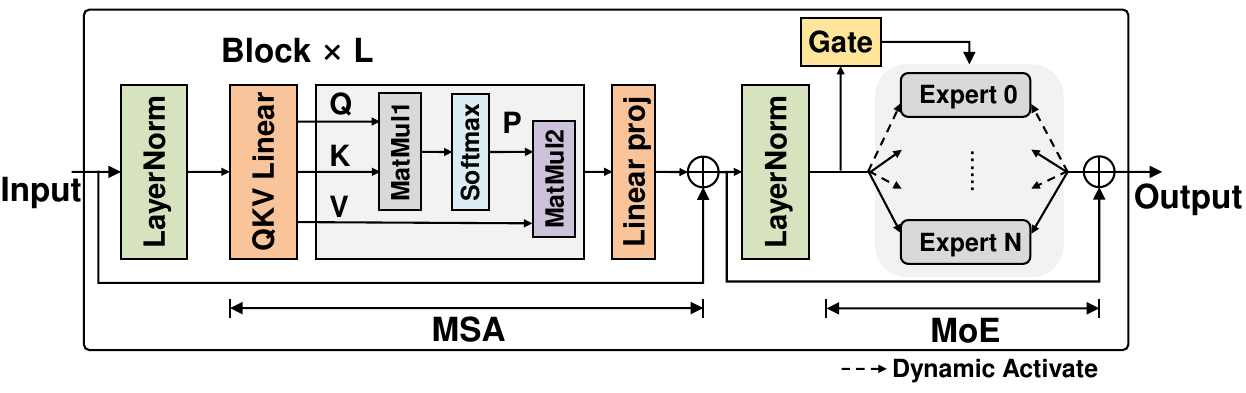}
\caption{The architecture of MoE-ViT that includes multiple
 Transformer blocks. Selected MLP blocks are replaced by MoE blocks.} \label{fig:VMoE}
 \vspace{-.5em}
\end{figure}
The Vision Transformer (ViT)\cite{dosovitskiy2020image} architecture processes input images by partitioning them into \( N \) flattened 2D patches, which are linearly projected into a sequence of \( D \)-dimensional embeddings \(\mathbf{X}_0 \in \mathbb{R}^{N \times D}\). These embeddings undergo hierarchical processing through \( L \) Transformer blocks, each containing a Multi-Head Self-Attention (MSA) module and a Multi-Layer Perceptron (MLP) module, with Layer Normalization (LayerNorm) and residual connections.



The MSA module performs the following computations on the activation $ X' $ obtained from the LayerNorm output:
\begin{equation}
\mathbf{Q}_i, \mathbf{K}_i, \mathbf{V}_i = \mathbf{X}' \mathbf{W}^{qkv}_i + \mathbf{b}^{qkv}_i
\end{equation}
\begin{equation}
\mathbf{P}_i = \text{Softmax}\left( \frac{\mathbf{Q}_i \mathbf{K}_i^\top}{\sqrt{D_h}} \right) ,\quad \mathbf{A}_i =\mathbf{P}_i\mathbf{V}_i ,\quad \text{MSA}(\mathbf{X}') = \mathbf{A} \mathbf{W}^o + \mathbf{b}^o
\label{attention}
\end{equation}
where $ \mathbf{W}^{qkv}_i \in \mathbb{R}^{D \times 3D_h} $, $ \mathbf{b}^{qkv}_i \in \mathbb{R}^{3D_h} $, $ \mathbf{W}^o \in \mathbb{R}^{hD_h \times D} $, $ \mathbf{b}^o \in \mathbb{R}^D $, $\mathbf{A} = [\mathbf{A}_0,\cdots \mathbf{A}_h ]$, $h$ is the number of the attention heads and $D_h$ is the feature size of each head.

The MLP module enhances feature expressivity through dimension expansion and non-linear activation, $\mathbf{Y}'$is the output obtained after applying LayerNorm to the output $Y$ of MSA:
\begin{equation}
\text{MLP}(\mathbf{Y}') = \text{GELU}(\mathbf{Y}' \mathbf{W}^1 + \mathbf{b}^1) \mathbf{W}^2 + \mathbf{b}^2
\end{equation}
where \( \mathbf{W}^1 \in \mathbb{R}^{D \times 4D} \)and \( \mathbf{W}^2 \in \mathbb{R}^{4D \times D} \).

MoE-ViT\cite{2021_vmoe, fan2022m3vit} extends this framework by replacing selected MLP layers with Mixture-of-Experts (MoE) blocks. As shown in Fig.~\ref{fig:VMoE}, each MoE block comprises \( m \) expert networks \(\{E_j\}_{j=1}^m\). A trainable gating network \( G: \mathbb{R}^D \to \mathbb{R}^m \) dynamically assigns weights to experts per input token \(\mathbf{y}_n \in \mathbf{Y}' \):
\begin{equation}
G(\mathbf{y}_n) = \text{Softmax}(\text{Top}_k(\mathbf{y}_n \mathbf{W}^g + \mathbf{b}^g))
\end{equation}
where \( \mathbf{W}^g \in \mathbb{R}^{D \times m} \), \( \mathbf{b}^g \in \mathbb{R}^m \), and \(\text{Top}_k\) retains the \(k\) largest values (typically \(k=1\) or \(2\)). The final output aggregates activated experts:
\begin{equation}
\text{MoE}(\mathbf{y}_n) = \sum_{j=1}^m G_j(\mathbf{y}_n) \cdot E_j(\mathbf{y}_n)
\end{equation}

This conditional computation induces sparsity by activating only task-relevant experts during inference. The architecture preserves ViT's global receptive field while enhancing model capacity without proportional increases in computational load.

\subsection{Model Quantization in ViT}

Model quantization \cite{li2023repq, lin2022fqvit, tai2023tsptq, shi2024p} has emerged as a prominent technique for neural network compression. This method operates by converting floating-point weights and activations into low-bit integer representations, thereby reducing memory requirements and computational overhead during inference. 

When deploying quantized models on FPGAs, hardware-friendly schemes are crucial. These utilize native operations like bit-shifting and lookup tables (LUTs), avoiding additional conversions. Uniform quantization is widely adopted for its hardware compatibility and efficient matrix multiplication, with two main variants:
\begin{itemize}
    \item \textbf{Asymmetric Quantization}: Uses zero-point offset $z$ to align tensor ranges
    \begin{equation}
        X_q = \left\lfloor X/s \right\rceil + z, \quad \hat{X} = s(X_q - z)
    \end{equation}
    
    \item \textbf{Symmetric Quantization}: Omits zero-point for simplicity
    \begin{equation}
        X_q = \left\lfloor X/s \right\rceil, \quad \hat{X} = sX_q
    \end{equation}
\end{itemize}

The computational advantage of symmetric quantization becomes evident when examining dequantization complexity. Consider matrix multiplication operations, asymmetric quantization induces a four-term expansion:
\begin{equation}
    XW = s_x s_w (X_q W_q - z_x W_q - z_w X_q + z_x z_w)
\end{equation}
whereas symmetric quantization reduces to a single product term:
\begin{equation}
    XW = s_x s_w X_q W_q
\end{equation}
This simplification significantly reduces computational overhead, though typically at the cost of greater precision degradation compared to asymmetric approaches.

\section{Quantization Scheme}
\begin{figure}[t]
\centering
\includegraphics[width=.9\textwidth]{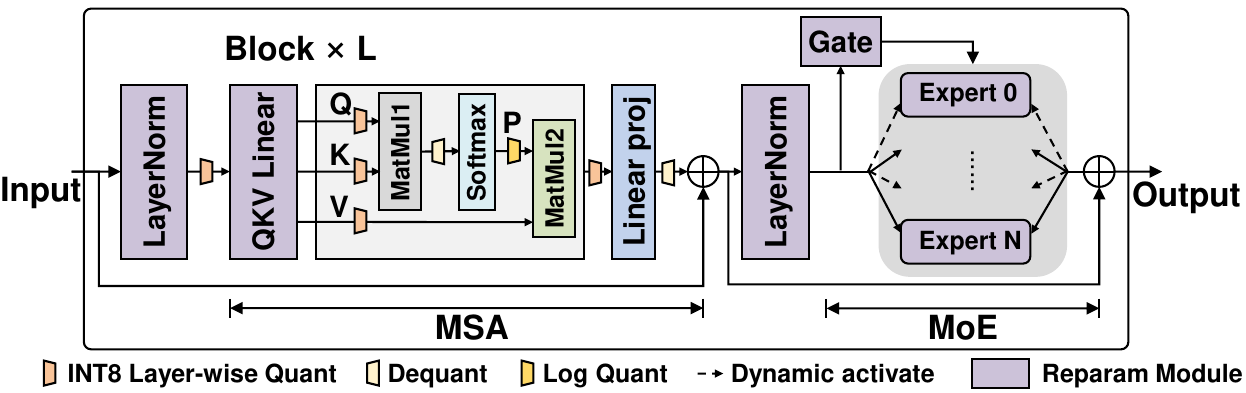}
\caption{Reparameterization-Based Quantization Scheme} \label{fig:Quant Scheme}
\vspace{-1.5em}
\end{figure}

To ensure quantization accuracy, we adopt customized quantizers tailored to the distribution characteristics of different activations. Reparameterization techniques are then employed to convert complex quantizers into hardware-friendly implementations. The entire quantization framework is illustrated in Fig.~\ref{fig:Quant Scheme}. Specialized quantization schemes are designed for the unique distributions of post-LayerNorm and post-Softmax activations, while INT8 symmetric quantization is applied to other linear layers.

\subsection{Reparameterization-Based Post-LayerNorm Quantization}  
\label{subsec:reparameterization}  

\begin{figure}[t]
\centering
\includegraphics[width=0.9\textwidth]{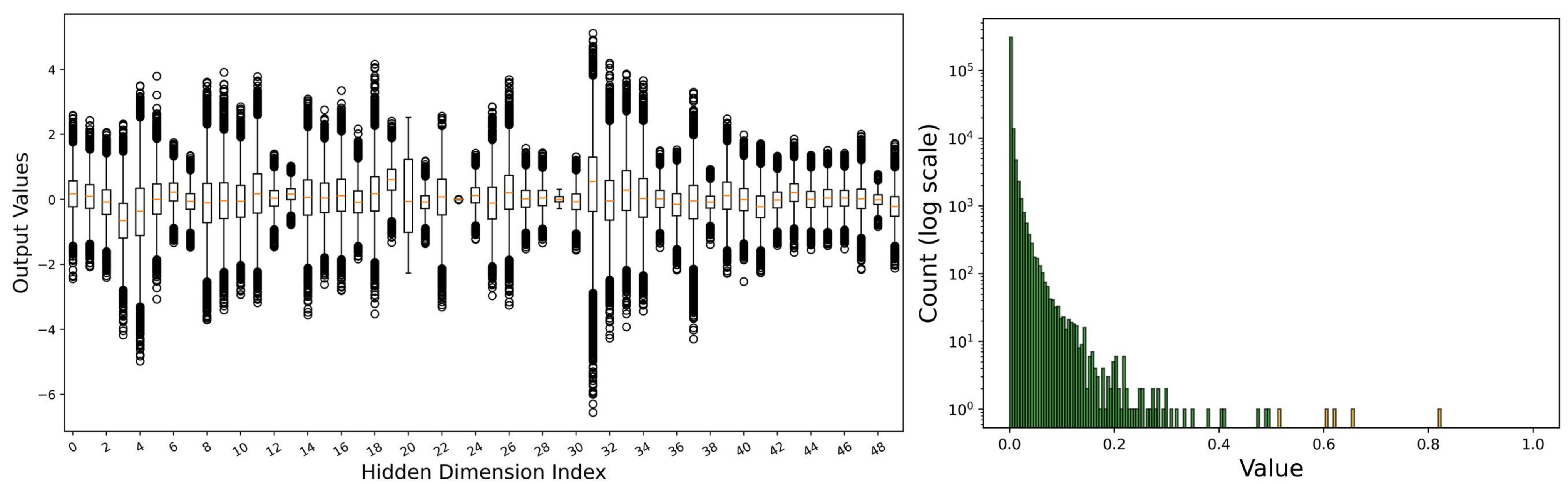}

\caption{Visualization of activation distributions from the first layer of ViT-Base.
Left: The post-LayerNorm activations exhibit strong inter-channel variance and asymmetric distribution.
Right: The post-Softmax activations follow a heavy-tailed distribution, with most values near zero and a few near one.}
\label{fig:activations}\
\vspace{-1.5em}
\end{figure}

As demonstrated in Figure \ref{fig:activations}, post-LayerNorm activations exhibit significant inter-channel variance and a non-symmetric distribution relative to zero. Although per-channel asymmetric quantization offers improved quantization error reduction, its reliance on dedicated hardware and increased computational overhead constrain its widespread application. To address this issue, we propose a reparameterization strategy that converts the per-channel asymmetric quantization of post-LayerNorm activations into a hardware-friendly per-layer symmetric quantization.

Given the LayerNorm output $\mathbf{X} \in \mathbb{R}^{N \times D}$, we first perform per-channel asymmetric quantization to obtain the channel-specific scale factors $\boldsymbol{s} \in \mathbb{R}^D$ and zero points $\boldsymbol{z} \in \mathbb{Z}^D$.
To convert these into the per-layer symmetric quantization scale factor $\tilde{s} = \mathbb{E}[\boldsymbol{s}]$, we define the transformation factors as
\begin{equation}
\boldsymbol{r_1} = \frac{\tilde{s}}{\boldsymbol{s}}, \quad \boldsymbol{r_2} = \boldsymbol{z} - 2^{b-1}\cdot\boldsymbol{1},
\end{equation}
where \(\boldsymbol{1}\) denotes a vector of ones. Reparameterizing the LayerNorm parameters adapts the distribution of \(\mathbf{X}\) to the unified scale factor \(\tilde{s}\) required for per-layer symmetric quantization.

Specifically, the LayerNorm parameters are updated via bias correction and parameter adjustment as follows:
\begin{equation}
\boldsymbol{\beta}' = \frac{\boldsymbol{\beta} + \boldsymbol{s} \odot \boldsymbol{r_2}}{\boldsymbol{r_1}}, \quad
\boldsymbol{\gamma}' = \frac{\boldsymbol{\gamma}}{\boldsymbol{r_1}},
\end{equation}
where \(\odot\) denotes the Hadamard product. Consequently, the transformed output is given by

\begin{equation}
\mathbf{X}' = \bigl(\mathbf{X} + \boldsymbol{s} \odot \boldsymbol{r_2}\bigr) \oslash \boldsymbol{r_1},
\end{equation}
with \(\oslash\) indicating element-wise division applied column-wise using the corresponding elements of \(\boldsymbol{r_1}\).

To preserve computational equivalence in the subsequent linear layer, an inverse transformation is applied to the weight matrix \(\mathbf{W}\) and bias vector \(\mathbf{b}\). Through equivalent transformations, we have
\begin{equation}
\mathbf{X}\mathbf{W} + \mathbf{b} = \mathbf{X}'\bigl(\operatorname{diag}(\boldsymbol{r_1}) \mathbf{W}\bigr) + \Bigl(\mathbf{b} - \mathbf{W}^T\bigl(\boldsymbol{s} \odot \boldsymbol{r_2}\bigr)\Bigr)
\end{equation}
Thus, to ensure that the output of the next linear layer remains invariant, we update the parameters as
\begin{equation}
\mathbf{W}' = \operatorname{diag}(\boldsymbol{r_1}) \mathbf{W}, \quad \mathbf{b}' = \mathbf{b} - \mathbf{W}^T\bigl(\boldsymbol{s} \odot \boldsymbol{r_2}\bigr)
\end{equation}

In the MoE architecture, a similar compensation mechanism is synchronously applied to the first linear layer of experts and the gating network for all experts:
\begin{equation}
{\mathbf{W}^{E_i}_{fc1}}' = \operatorname{diag}(\boldsymbol{r_1}) \mathbf{W}^{E_i}_{fc1}, \quad {\mathbf{b}^{E_i}_{fc1}}' = \mathbf{b}^{E_i}_{fc1} - \Bigl({\mathbf{W}^{E_i}_{fc1}}\Bigr)^T\bigl(\boldsymbol{s} \odot \boldsymbol{r_2}\bigr)
\end{equation}
\begin{equation}
\mathbf{W}_{gate}' = \operatorname{diag}(\boldsymbol{r_1}) \mathbf{W}_{gate}, \quad \mathbf{b}_{gate}' = \mathbf{b}_{gate} - \mathbf{W}_{gate}^T\bigl(\boldsymbol{s} \odot \boldsymbol{r_2}\bigr)
\end{equation}

Analogous adjustments are applied to the parameters \(\mathbf{W}_{fc1}\) in the MLP and \(\mathbf{W}_{qkv}\) in MSA.

Through this reparameterization strategy, we transform per-channel asymmetric quantization into a hardware-friendly per-layer symmetric quantization scheme, thereby retaining the inference acceleration benefits of per-layer symmetric quantization with only a marginal degradation in performance compared to per-channel asymmetric quantization.

\subsection{Reparameterization-Based Post-Softmax Quantization}  
\label{subsec:post_softmax_quant}  

In vision transformers, Softmax operations map attention scores to probability distributions within $(0,1)$. As shown in Fig.~\ref{fig:activations}, post-Softmax activations exhibit heavy-tailed distributions where over 99\% of values are below 0.3, while the remaining 1\% near 1.0 carry critical semantic information. Conventional clipping strategies fail to preserve these high-value components.

The non-uniform quantizer effectively quantizes post-softmax activations. We define the \(\log_{\sqrt{2}}\) quantizer as follows:  

\begin{equation}
    \mathbf{A}_q = \text{clip}\left(\left\lfloor -\log_{\sqrt{2}} \mathbf{A} \right\rceil, 0, 2^b - 1\right)
\end{equation}

To facilitate hardware implementation, we optimize the softmax computation process. Specifically, we apply quantization to the numerator in the formulation, which has a range of \((0,1)\). Consequently, the scaling factor $\boldsymbol{s}$ is set to 1.  

Compared to the \(\log_2\) quantizer, the \(\log_{\sqrt{2}}\) quantizer offers higher precision but is hardware-unfriendly due to complex logarithmic computations. We reparameterize it into a more hardware-efficient form.
\begin{align}
    \mathbf{A}_q &= \text{clip}\left(\left\lfloor -2 \log_2 \mathbf{A} \right\rceil, 0, 2^b - 1\right) \\
    \widehat{\mathbf{A}} &= 2^{-\frac{\mathbf{A}_q}{2}} = 
    \begin{cases} 
    2^{-k}, & \mathbf{A}_q = 2k \\
    2^{-(k+1)}, & \mathbf{A}_q = 2k + 1
    \end{cases}
    \quad \text{for } k \in \mathbb{Z} \nonumber \\
    &= 2^{\left\lfloor -\frac{\mathbf{A}^{(Z)}}{2} \right\rfloor} \cdot \left[ \mathbbm{1}(\mathbf{A}^{(Z)}) \cdot (\sqrt{2} - 1) + 1 \right] 
\end{align}
where $\mathbbm{1}(\cdot)$is a parity indicator function that is 0 at even numbers and 1 at odd numbers.

Consequently, we derive the new quantization scaling factor:  
\begin{equation}
    \boldsymbol{s}' = 1 \cdot \left[ \mathbbm{1}(\mathbf{A}_q) \cdot (\sqrt{2} - 1) + 1 \right]
\end{equation}

This formulation preserves numerical precision while enhancing compatibility with hardware-efficient implementations.
This allows efficient attention-value matrix multiplication using bit-shift operations:
\begin{equation}
    \mathbf{A}\cdot \mathbf{V}_q = \left(\mathbf{V}_q \gg \left\lfloor \frac{\mathbf{A}_q}{2} \right\rfloor \right) \cdot \boldsymbol{s}'
\end{equation}
where $\gg$ denotes right bit-shift and the scaling factor $\boldsymbol{s}'$ will be fused into $\mathbf{V}_q$'s quantization parameters.
\section{Hardware Design}

To fully realize the algorithmic advantages of our algorithm, developing a dedicated accelerator for quantized MoE-ViT is essential. However, this effort faces significant challenges, particularly with memory bandwidth limitations and hardware resource inefficiency compared to standard ViTs. To overcome these obstacles, we decoupled computation from I/O operations and restructured the computation order at multiple granularities to reduce bandwidth pressure. Additionally, we further fused the quantized softmax computation process, leading to improved efficiency in both resource utilization and performance.

\subsection{Overall Architecture}
\begin{figure}[t]
\centering
\includegraphics[width=.78\textwidth]{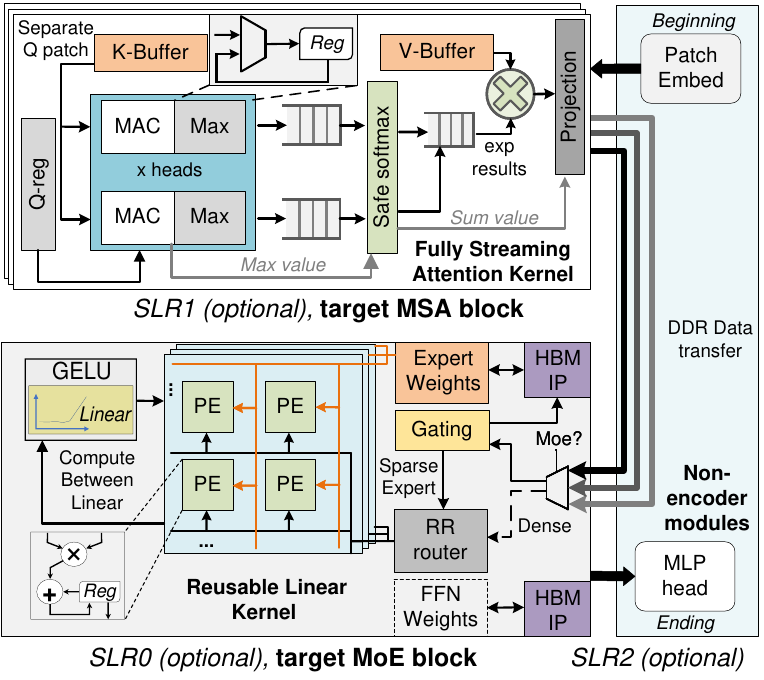}
\caption{The overall architecture of \placeh.} \label{fig:overall2}
\vspace{-1.5em}
\end{figure}

Fig.~\ref{fig:overall2} illustrates the architecture of \placeh. Our work primarily targets the matrix multiplication modules, with a particular focus on attention computation and linear transformations. 
Host-device communication occurs only at the beginning and end of execution. Specifically, input data is transferred from host memory to the FPGA’s off-chip memory (e.g., DDR or HBM) using OpenCL buffer operations before kernel launch. The execution of FPGA kernels is also managed via OpenCL command queues, which schedule and dispatch the tasks to the device. During kernel execution, both MSA and MoE blocks are performed independently on the FPGA by accessing only on-chip or off-chip memory, without further interaction with the host. After execution completes, the results are transferred back to host memory. 

While full on-chip caching offers significant acceleration benefits for compact models such as ViT-Tiny and ViT-Small, it becomes impractical for large-scale MoE architectures due to their dynamic and extensive runtime memory demands. To overcome this limitation, our design enables flexible memory management across both on-chip and off-chip resources. Specifically, the attention and linear kernels support dual-mode data access: activations and weights can either be preloaded into on-chip buffers to maximize data reuse, or streamed directly from off-chip memory during execution to accommodate larger model sizes. This hybrid strategy ensures adaptability to varying model scales and aligns with the memory hierarchy and bandwidth characteristics of different FPGA platforms.


\subsection{Bandwidth Optimization in Parallel Computation}
Quantization can effectively reduce computational resource usage. However, simply increasing parallelism to improve resource utilization does not always yield linear performance gains, as memory bandwidth often becomes a critical bottleneck, limiting system efficiency and scalability. This issue is particularly pronounced in MoE architectures, where frequent off-chip loading of expert weights exacerbates memory access demands. To address this challenge, we redesigned the memory access patterns of both the attention and linear kernels. 

\subsubsection{Coarse-grained Patch Reorder in attention}

\begin{figure}[t]
\centering
\includegraphics[width=\textwidth]{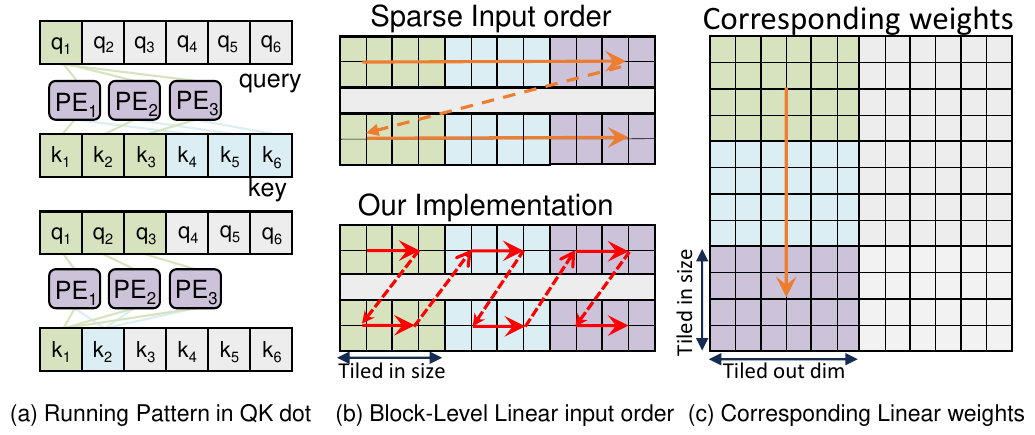}
\caption{Memory Access Patterns in Attention Mechanisms and Linear Kernels.} \label{fig:reorder}
\vspace{-2em}
\end{figure}

In conventional softmax computation, the naive blockwise approach requires each processing element (PE) to sequentially compute partial key vectors \( K_j \)  and transfer them to dedicated softmax units. This leads to significant off-chip memory overhead, as each transaction fetches \(N_{PE}\) data blocks—an especially inefficient design for quantized implementations, where higher DSP efficiency demands more PEs.

This architectural consideration motivates our innovative design shown in Fig.~\ref{fig:reorder}(a): Rather than distributing distinct \(K\) values across PEs, we implement a broadcast mechanism that propagates identical \(K\) tensors to all PEs, while uniquely assigning query vectors \( Q_i \) to corresponding PE. This paradigm shift achieves critical advantages, as the number of off-chip memory accesses remains constant regardless of PE scaling.
\subsubsection{Fine-grained Linear Control Logic}
Fig.~\ref{fig:reorder}(b) illustrates the execution workflow, revealing a critical inefficiency in conventional single-CU (Compute Unit) architectures: Each patch computation requires full-weight re-fetching regardless of sparsity patterns. As model dimensions scale exponentially, this process incurs non-linear growth in data transfer overhead, fundamentally constrained by memory bandwidth limitations.

To overcome these challenges, we have developed a unified linear kernel with multi-scenario adaptability. We deploy \( N_{\text{L}} \) CUs in parallel, with a partitioned activation prefetching mechanism through a single round-robin (RR) router. During execution, the hardware router dynamically selects the first \( N_{\text{L}} \) available patch indices, generates corresponding mapping addresses, and streams tiled vectors to multiple CUs through cyclic distribution.

Compared to directly using multiple linear kernels, the selection policy abstraction in our design enables transitions between sparse/dense computation modes through runtime-reconfigurable parameters.
Also, due to the temporal locality of weight references (Fig.~\ref{fig:reorder}(c)), our approach reduces off-chip memory access pressure during runtime, making it suitable for deploying larger-scale models.

\subsection{Hardware-Aware Parallel Softmax Acceleration}

Benefiting from the previously modified computation in the attention mechanism, each PE performs complete computations without inter-PE partial result exchanges. This parallelism significantly simplifies the computational workflow, thereby inspiring us to implement a fully on-chip fused softmax module.

For simplicity, we use the traditional 3-pass softmax as the base for our explanation. 
In \textbf{Pass 1}, since each PE retains a distinct Q, the maximum value extraction is performed within the pipeline using a single max module, eliminating the need for additional termination logic. When Q is updated, the maximum value is directly propagated to the next stage.

In \textbf{Pass 2}, PE2 receives the maximum value and the L buffer to compute \( f(x) \) and accumulates \( l(x) \). Our hardware algorithm directly processes the softmax numerator rather than the final result, allowing \( f(x) \) to be immediately passed to the next CU for multiplication without introducing latency. 

In \textbf{Pass 3}, multiplication with V is implemented using shift operations instead of traditional multipliers, further reducing computational overhead. Since all rows share the same denominator, the final result is obtained by multiplying the output with \(\text{recip}(l(x)) \cdot s_v\). This step requires only \( T_s \) multipliers, ensuring efficient resource utilization.

\section{Experiments} \label{sec:experiments}
Our contribution primarily targets MoE-ViT, but the lack of existing quantized MoE accelerator implementations makes our validation relatively limited. Thus, we also conducted experiments and comparisons within ViT-related work to validate the feasibility of our results.

\subsection{Experimental Setup}
\subsubsection*{Quantization Evaluation} We evaluate ViTs and DeiTs quantization performance on ImageNet\cite{deng2009imagenet}. As for quantization details, we randomly select 32 images from ImageNet's training set as the calibration data, which are used to analyze activation distributions for offline calculating scaling factors of activations and weights\cite{lin2022fqvit, liu2021post, xiao2023smoothquant}, and then evaluate accuracy on its validation set. 

\subsubsection*{Hardware Deployment and Platform Selection} 
We deploy MoE-ViT on the Xilinx ZCU102 and Alveo U280 platforms using Vitis HLS and Vivado (v2022.2). Since VMoE~\cite{2021_vmoe} lacks a PyTorch implementation, we adopt $\text{M}^3$ViT~\cite{fan2022m3vit}, which shares the same computation but differs in execution order. We use PyTorch (v2.0.1) with a batch size of 4. 

The ZCU102 features a single-SLR architecture with limited resources (2520 DSPs, 1824 BRAMs, 21 GB/s DDR bandwidth). In contrast, the U280 provides a multi-SLR architecture with significantly more resources (9024 DSPs, 2160 BRAMs distributed across 3 SLRs) and high memory bandwidth (460 GB/s) via 8 GB of HBM2. In our evaluation, latency specifically denotes the actual kernel execution time, defined as the interval between enqueueing the kernel command and receiving the completion signal.

\begin{table}[t]
    \centering
    \setlength{\tabcolsep}{4pt}
    \caption{Quantization results of image classification on ImageNet dataset, where each data presents the Top-1 accuracy (\%).``W/A/Attn'' indicates that the quantization bit-width of weights/activations/attention maps, respectively.}
    \renewcommand{\arraystretch}{1.2}
    \resizebox{1.0\linewidth}{!}{
    \begin{threeparttable}{
    \begin{tabular}{c|cccccccc} \Xhline{3\arrayrulewidth}
         \textbf{}  & \textbf{W/A/Attn} & \textbf{M\(^3\)ViT} & \textbf{ViT-T} & \textbf{ViT-S} & \textbf{ViT-B} & \textbf{DeiT-T} & \textbf{DeiT-S} & \textbf{DeiT-B}\\ \hline \hline
         \textbf{Full Precision} & 32/32/32 & 85.17 & 75.46 & 81.39  & 84.53 & 72.21 & 79.85 & 81.85\\ \hline
         \textbf{MinMax\cite{lin2022fqvit}} & 8/8/8 & 82.54 & 19.87 & 30.28  & 23.64 & 70.94 &  75.05 & 78.02\\ \hline
         \textbf{RePQ-ViT\cite{li2023repq}}$^*$ & 8/8/8 & - & 72.85 & 81.31  & 84.36 & 71.94 &  79.76 & 81.79 \\ \hline
         \textbf{TSPTQ-ViT\cite{tai2023tsptq}} & 8/8/8 & - & - & 81.20  & 84.11 & 71.87 &  79.56 & 81.72 \\ \hline
         \textbf{FQ-ViT\cite{lin2022fqvit}} & 8/8/4 & - & - & -  & 82.68 & 71.07 & 78.40 & 80.85\\ \hline
         \textbf{P2-ViT\cite{shi2024p}} & 8/8/4 & - & - & - & 82.80 & 70.92 & 78.24 & 80.96\\ \hline
         \textbf{Our Method} & 8/8/4 & 84.89 & 71.79 & 80.12 & 83.99 & 71.64 & 79.38 & 81.60\\ \Xhline{3\arrayrulewidth}
    \end{tabular}}
    \captionsetup{justification=raggedright,singlelinecheck=false}
    \caption*{\footnotesize $^*$ Reproduced from open-source code.}
    \end{threeparttable}}
    \vspace{-2.5em}
    \label{table:quantization}
\end{table}
\begin{table}[t]
    \centering
    \footnotesize
    \setlength{\tabcolsep}{5.5pt}
    \caption{Resource consumption of deploying \placeh~on ZCU102 and U280.}
    \renewcommand{\arraystretch}{1.2}
    \resizebox{0.98\linewidth}{!}{
    \begin{threeparttable}{
    \begin{tabular}{c|cccccc} \Xhline{3\arrayrulewidth}
          \textbf{Platform} & \textbf{Model}  & \textbf{DSPs} & \textbf{BRAMs} & \textbf{LUTs} & \textbf{Flip-Flop (FFs)} & \textbf{Power} \\ \hline \hline
         \textbf{ZCU102} & \textbf{ViT-tiny} & 1754 & 383 & 156.1K & 198.2K  & 9.83W \\ \hline
         \textbf{U280} &  \textbf{ViT-small} & 2635 & 696.5 & 311.6K & 454.1K & 33.7W \\ \Xhline{3\arrayrulewidth}
    \end{tabular}}
    \end{threeparttable}}
    \vspace{-1.5em}
    \label{table:Resource}
\end{table}
\subsection{Quantization Algorithm Validation}
To validate the effectiveness of the proposed method, we conducted experiments on the ImageNet dataset using various architectures, including ViT \cite{dosovitskiy2020image} and DeiT \cite{deit}, and compared the results with previous works\cite{li2023repq, tai2023tsptq, lin2022fqvit, shi2024p}. As summarized in Table~\ref{table:quantization}, our approach achieves hardware-compatible quantization while maintaining minimal accuracy degradation. Notably, the method delivers 83.99\% top-1 accuracy under an 8/8/4-bit configuration on ViT-B. In addition, the accuracy loss of our quantization scheme on M$^3$ViT is also presented in Table~\ref{table:quantization}, showing only a 0.28\% reduction compared to full precision.

\subsection{Comparison With Prior Transformer Accelerators}
Since we employed a unified kernel to ensure compatibility for executing both MoE and MLP layers, our accelerator can support both MoE models and standard ViT simply by switching the operation mode. Given that the same precision was used throughout, the optimal configurations for both components in the experiments were essentially identical. Consequently, we present the resource information in Table~\ref{table:Resource}. 

Given that our kernel design is decoupled from parallelism, the utilization of computational resources becomes the primary bottleneck. For the ZCU102 platform, we manually connected the IP cores in Vivado, which allowed for finer control over resource allocation and led to improved hardware utilization.

Despite U280's rich resources, the Vitis flow’s automatic placement, routing, and IP insertion can lead to congestion in multi-SLR designs. We therefore limit usage to two SLRs and balance their latency to improve frequency.

\begin{table}[tbp]
\footnotesize
\centering \vspace{-1em}
\setlength{\tabcolsep}{12pt}
\caption{Comparison with GPU and Edge-MoE\cite{sarkar2023edge} on $\text{M}^3$ViT} 
\renewcommand{\arraystretch}{1.2}
\resizebox{0.98\linewidth}{!}{
\begin{threeparttable}{
\begin{tabular}{l|c|c|c|c|c}
\Xhline{3\arrayrulewidth}
\textbf{Attribute} & \multicolumn{2}{c|}{\textbf{M\(^3\)ViT}} & \textbf{Edge-MoE} & \multicolumn{2}{c}{\textbf{Ours} } \\ \hline \hline

\textbf{Platform} & \multicolumn{2}{c|}{Tesla V100S} & ZCU102 & ZCU102 & U280 \\ \hline
\textbf{Bit-width} & \multicolumn{2}{c|}{FP32} & $W^{16}A^{32}$ & INT8 & INT8 \\ \hline
\textbf{Model} & $\text{M}^3$ViT-T & $\text{M}^3$ViT-S & $\text{M}^3$ViT-T & $\text{M}^3$ViT-T & $\text{M}^3$ViT-S\\ \hline

\textbf{Frequency (Mhz)} & 1245 & 1245 & 300 & 300 & 250 \\ \hline
\textbf{Power (W)} & 42.98 & 47.12 & 14.54 & 9.83 & 33.7 \\ \hline
\textbf{Latency (ms)} & 3.65 & 4.45 & 34.64 & 6.47 & 9.16 \\ \hline
\textbf{Throughput (GOPS)} & 561.79 & 1936.84 & 72.15 & 386.3 & 1004.3 \\ \hline
\textbf{Efficiency (GOPS/W)} & 11.88 & 19.25 & 4.83 & \textbf{38.639} & \textbf{29.8} \\ 
\Xhline{3\arrayrulewidth}
\end{tabular}}
\end{threeparttable}}
\label{table:M3vit} \vspace{-1.1em}
\end{table}

\subsubsection*{Comparisons with M$^3$ViT Baselines on GPUs and Edge-MoE}
As shown in Table~\ref{table:M3vit}, our \placeh's accelerators achieve 68.7\% and 51.8\% of the GPU throughput, respectively, with absolute values of 386.3 GOPS and 1004.3 GOPS. This gap is expected, as FPGAs inherently possess fewer computational and memory resources than GPUs.

In contrast, \placeh~demonstrates significant advantages in energy efficiency, achieving 38.639 GOPS/W and 29.8 GOPS/W, which correspond to 3.25$\times$ and 1.54$\times$ improvements over the GPU. Compared with Edge-MoE on the ZCU102 platform, \placeh~achieves a 7.99$\times$ gain in energy efficiency. We attribute these improvements primarily to the hardware-friendly quantization scheme, which introduces negligible overhead for both quantization and dequantization. 

\subsubsection*{Comparisons with FPGA-based Accelerators}
To facilitate comparison with prior works, we implemented the proposed \placeh~accelerators on FPGA platforms. The primary comparison metric is GOPS/W, highlighting energy efficiency improvements over conventional ViT implementations. 

Compared to M$^3$ViT, the standard ViT implementation adopts pre-loaded weights to further reduce latency. Specifically, by eliminating runtime off-chip data access and leveraging double-buffering across kernels, the design achieves improved pipelining and reduced end-to-end latency. As shown in Table~\ref{table:vit compare}, in terms of latency, \placeh~can obtain 1.65$\times$ and 1.33$\times$ acceleration compared to HeatViT. Also, we can obtain a large energy efficiency improvement. Specifically, \placeh-E attains a 1.11$\times$ enhancement over the HeatViT, while \placeh-C attains 1.71$\times$ and 1.89$\times$ improvement when compared with the TECS'23\cite{ye2023accelerating} and ASP-DAC'24\cite{aspdac24_swat}.

\begin{table}[tbp] 
\setlength{\tabcolsep}{1.pt}
 \caption{Comparison with Previous FPGA Implementations} 
 \renewcommand\arraystretch{1.2}
\resizebox{\linewidth}{!}{
\begin{tabular}{l|c|c|c|c|c|c} 
\Xhline{3\arrayrulewidth}

\textbf{Attribute} & \textbf{ViA~\cite{wang2022via}}& \textbf{HeatViT~\cite{dong2023heatvit}} & \textbf{\placeh-E} & \textbf{TECS'23 ~\cite{ye2023accelerating}} &\textbf{ASP-DAC'24~\cite{aspdac24_swat}} & \textbf{\placeh-C}\\
\hline \hline
\textbf{Model} & Swin-T & DeiT-S & ViT-T & BERT-B & Swin-T & ViT-S \\
\hline
\textbf{Platform} & U50 & ZCU102 & ZCU102 & U250 & U280 & U280 \\
\hline
\textbf{Bit-width} & FP16 & INT8 & INT8 & INT8 & INT8 & INT8 \\
\hline
\textbf{Freq. (MHz)} & 300 & 300  & 300  & 300 & 200 & 250  \\ \hline
\textbf{Power (W)} & 29.6 & 10.697  & 9.83  & 77.168  & 32.24 & 33.7  \\
\hline
\textbf{Latency (ms)} & - & 9.15 & 5.53 & - & - & 6.84 \\
\hline
\textbf{Thpt. (GOPS)} & 309.6  & 440.0  & \textbf{452.08} & 1800 & 309.1 & \textbf{1345.0}  \\ \hline
\textbf{Effi. (GOPS/W)} & 7.94 & 41.12 & \textbf{45.98} & 23.32 & 21.04 & \textbf{39.91} \\
\Xhline{3\arrayrulewidth}
\end{tabular}} \label{table:vit compare} 
\vspace{-2em}
\end{table} 

\section{Conclusion}
In this paper, we introduce \placeh, the first end-to-end FPGA implementation for quantized MoE-ViT. By combining hardware-friendly algorithms with a design approach that incorporates multiple levels of granularity, our approach achieves low latency and high resource efficiency. On the U280 platform, it delivers M$^3$ViT-small processing at 109 FPS, equivalent to 1004.3 GOPS, outperforming previous state-of-the-art MoE-ViT accelerators.

\subsubsection*{Acknowledgments}
This work was supported in part by the Strategic Priority Research Program of the Chinese Academy of Sciences, Grant No. XDB0660101, XDB0660000, and XDB0660100, in part by the National Natural Science Foundation of China under Grant 62172380, in part by Jiangsu Provincial Natural Science Foundation under Grant BK20241818, in part by USTC Research Funds of the Double First-Class Initiative under Grant YD2150002011.

\subsubsection*{Disclosure of Interests}
The authors have no competing interests to declare that are relevant to the content of this article.

%
%
%
\bibliographystyle{splncs04}
\bibliography{refs_simplified_fixed}
%



\end{document}